\documentclass[structabstract]{aa} 
\usepackage{graphicx}
\usepackage{txfonts}
\usepackage[round]{natbib}
\bibpunct{(}{)}{;}{a}{}{,}
\DeclareMathSymbol{\lesssim}{\mathrel}{AMSa}{"2E}
\begin{document}
\bibliographystyle{aa}
 \title{The  long-term evolution of  the X-ray pulsar XTE J1814-338:  a receding jet
contribution to the quiescent optical emission?\thanks{Based on observations made with ESO Telescopes at the Paranal Observatory under programme ID 383.D-0730(A)}}

   \author{M. C. Baglio 
          \inst{1, 2}          
          ,\         
          P. D'Avanzo \inst{1}
          ,\
          T. Mu\~{n}oz-Darias \inst{3}
          ,\
          R. P. Breton \inst{3}
          \and
          S. Campana \inst{1}
          }
		  
   \institute{INAF, Osservatorio Astronomico di Brera, via E. Bianchi 46, I-23807 Merate (Lc), Italy\\
              \email{cristina.baglio@brera.inaf.it}
         \
            \\ 
            \and
            Universit\`{a} degli Studi di Milano, Dipartimento di Fisica, Via Celoria 16, I-20133 Milano, Italy
             \\ 
             \and
             School of Physics and Astronomy, University of Southampton, SO17 1BJ, UK
             }             

   \date{ }

   \abstract
   {}
   {We present a study of the quiescent optical counterpart of the Accreting Millisecond X-ray Pulsar XTE J1814-338 aimed at unveiling the different components contributing to the quiescent optical emission of the system.}
   { We carried out multiband ($BVR$) orbital phase-resolved photometry of the system using the ESO Very Large Telescope (VLT) equipped with the FORS2 camera, covering about $70 \%$ of the known 4.3 hour orbital period.}
   {The optical light curves are consistent with a sinusoidal variability modulated with the orbital period with a 0.5-0.7 mag semi-amplitude. They show evidence for a strongly irradiated companion star, in agreement with previous findings for this system. However, the observed colours cannot be accounted for by the companion star alone, suggesting the presence of an accretion disc during quiescence. The system seems to be fainter in all analysed bands compared to previous observations. The $ R $ band light curve displays a possible phase offset with respect to the $ B $ and $ V $ band. Through a combined fit of the multi-band light curve performed with a Markov chain Monte Carlo technique we derive constraints on the companion star and disc fluxes, on the system distance and on the companion star mass.}
   {The irradiation luminosity required to account for the observed day-side temperature of the companion star is consistent with the spin-down luminosity of  a millisecond radio pulsar. The flux decrease and spectral evolution of the quiescent optical emission observed comparing our data with previous observations, collected over 5 years, cannot be satisfactorily explained with the combined contribution of an irradiated companion star and of an accretion disc alone. The observed progressive flux decrease as the system gets bluer could be due to a continuum component evolving towards a lower, bluer spectrum. While most of the continuum component is likely due to the disc, we do not expect it to become bluer in quiescence. Hence we hypothesize that an additional component, such as synchrotron emission from a jet was contributing significantly in the earlier data obtained during quiescence and then progressively fading or moving its break frequency toward longer wavelengths.}

\authorrunning{Baglio, D'Avanzo, Munoz-Darias, Breton \& Campana} 
\titlerunning{Variability in the quiescent optical emission of XTE J1814-338}
\maketitle

\section{Introduction}

Fifteen Accreting Millisecond X-Ray Pulsars (AMXPs) have been discovered so far. These transient systems are a subclass of Low Mass X-Ray Binaries (LMXBs), hosting a weakly magnetic neutron star which has been spun up to millisecond periods by angular momentum transfer from a donor companion of mass $ M\lesssim 1M_{\odot} $.
These systems have an orbital period in the range of 40 min to 19 hr and spin frequencies between 1.7 and 5.4 ms (see \citealt{Patruno} for a recent review).

Such systems are suspected to be the ``missing link" between LMXBs and millisecond radio pulsars, the former being the progenitors of the latter. According to the standard model for the formation of millisecond radio pulsars, these objects come from the recycling of neutron stars hosted in LMXBs, i.e. have been spun up by the accretion torques resulting from the transfer of mass with a large angular momentum from the companion star. The radio emission switches off during the active mass transfer phase and resumes again when it comes to an end, leaving behind a recycled radio pulsar which looses rotational energy in the form of a relativistic particle wind that can irradiate the companion star.

The first confirmation of such a scenario has come with the observation of PSR J1023+0038,  also called ``missing link pulsar'' \citep{Archibald2009}, showing that LMXBs can turn on as radio millisecond pulsars. Furthermore, the recent detection of transient millisecond X-ray pulsations from the system IGR J18245-2452, which was known to be a millisecond radio pulsar, provided the direct observational evidence for the recycling scenario \citep{Papitto2013}.

Optical observations of AMXPs in quiescence offer the only chance of studying the characteristics of the companion star, otherwise veiled by the bright disc emission during active phases. In LMXBs comprising a companion star suffering significant tidal distortion, the light curves in quiescence are characterized by orbital variability due to ellipsoidal modulation. Such an effect arises from the fact that the projected surface area of the distorted star is larger at quadratures than at conjunctions, hence resulting in maxima and minima of the total flux at those two sets of phases, respectively. Gravity darkening introduces a slight asymmetry in the double modulation, making the minimum at superior conjunction of the companion deeper as a result of the lower surface gravity. However, in systems with short orbital periods (and hence with a smaller orbital separation) the companion star might be severely irradiated by the compact object and/or by the residual accretion disc. In such cases, the light curve would be dominated by the contribution from the irradiation effect, which displays a maximum at superior conjunction of the companion and a minimum at inferior conjunction. Additional features can be observed in the light curve profiles, particularly if a residual accretion disc is significantly contributing to the overall optical emission during quiescence (see, e.g., \citealt{Hynes2010}). Detailed multiband optical and near-infrared orbital phase resolved monitoring of these systems can provide useful information to disentangle the different components contributing to their total quiescent emission.

The binary system XTE J1814-338 was discovered during an outburst on 2003 June 5, MJD 52795, by the {\it Rossi X-ray Timing Explorer} ({\it RXTE}) (\citealt{MarkwardtSwank}). The system is characterized by coherent pulsations at a frequency of 314.3 Hz (3.18 ms),  with an orbital modulation of 4.3 hr. Thanks to the observation of type I X-ray bursts, \citet{Strohmayer} estimated the distance of this system to be $8.0 \pm 1.6 \, \rm kpc$. During the outburst an $R\sim18.3$ optical counterpart was identified and hydrogen and helium emission lines were detected through optical spectroscopy, suggesting for a non-degenerate companion \citep{Krauss2005}.
The quiescent optical counterpart observed in 2004 is rather faint ($R \sim 22.5$ and $V \sim 23.3$; \citealt{Davanzo}). The orbital phase-resolved quiescent optical light curves display a sinusoidal profile showing a maximum at the superior conjunction of the companion. Such a behaviour suggests that irradiation from the compact object plays a key role at heating the inner face of the companion star in the observed optical quiescent luminosity. This has been intepreted as the indirect signature of an energetic rotation-powered recycled millisecond pulsar (\citealt{Davanzo} and references therein).

In this paper we present ESO-VLT orbital phase-resolved multiband optical imaging observations of XTE J1814-338 obtained on September 2009, together with archival data obtained in May 2004. These observations are complementary to those presented in \citet{Davanzo}. These datasets were obtained while the source was in quiescence and cover a period of five years. No outburst occurred between the two epochs. 

The paper is organized as follows: in Section 2 we present our data, in Section 3 we compare our results with the findings of \citet{Davanzo} and present a discussion on the origin of the quiescent optical emission of XTE J1814-338. Our conclusions are reported in Section 4 and 5. Through the paper all errors are at $ 68 \% $ confidence level unless stated differently.

\section{Observations and data analysis}
  
%

XTE J1814-338 was observed in quiescence on 10 September 2009 with the ESO Very Large Telescope (VLT), using the FORS2 camera in imaging mode with the \textit{B, V, R} filters. The night was clear, with seeing in the range $0.6''-2.5''$ degrading through the night. A set of 29 images of 3 minutes integration each was obtained for each filter, covering more than one orbital period. However, due to the crowdness of the field, only images obtained under seeing $ \lesssim 1'' $ had a quality good enough for our purposes. We thus retain 19 images for the $B$ and $V$ filters, and 16 for the $R$ one, that cover about $70 \%$ of the system's orbital period.
   
We also retrieved from the ESO Archive $ I $ band FORS2 VLT imaging data obtained on 2004 May 21 and 22, which are contemporary with the $ R $ and $ V $ data reported in \citet{Davanzo}.
   
Image reduction was carried out following standard procedures: subtraction of an averaged bias frame and division by a normalized flat frame. Astrometry was performed using the
USNOB1.0\footnote{http://www.nofs.navy.mil/data/fchpix/} catalogue.
Point Spread Function-photometry was made with the
ESO-MIDAS\footnote{http://www.eso.org/projects/esomidas/} 
daophot\footnote{http://http://www.star.bris.ac.uk/~mbt/daophot/} task for
all the objects in the field. The photometric calibration was done against the Stetson standard field star PG2213 \citep{Stetson}. In order to minimize any systematic effect, we
performed differential photometry with respect to a selection of 8 local
isolated and non-saturated reference stars, which
were also used as secondary standards to cross-check the calibration of the
2004 and 2009 data in the $V$ and $R$ band.

\section{Results}
\subsection{2004 $ I $ band data}
\begin{figure}
\begin{center}
\includegraphics[scale=0.25]{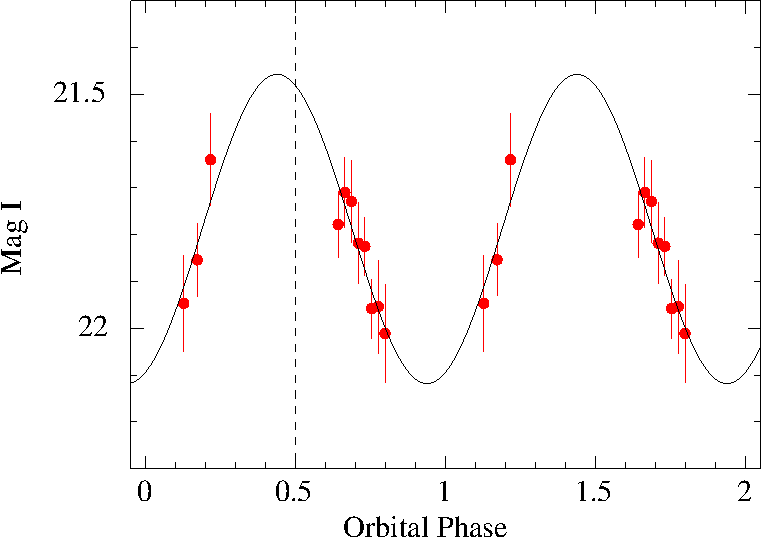}
      
   \caption{$I$ band light curve for the system XTE J1814-338 for the 2004 observation. Phase 0.0 corresponds to the inferior conjunction of the companion. We calculated
             the phases based on the X-rays ephemeridis of \citet{Papitto}. Two periods of the system are drawn for clarity.}
\label{lc_I}
\end{center}
  
   \end{figure}

The 2004 $ I $ band light curve shows variability consistent with a sinusoid modulated at the 4.3-hr orbital period, with a semi-amplitude of 0.33 mag and an average magnitude of $I = 21.79 \pm 0.04$ (Fig. \ref{lc_I}; Tab \ref{phot}). The light curve shows a single minimum around phase 0.0 (inferior conjunction of the companion) and a maximum around 0.5 (superior conjunction of the companion), overall consistent with the $ V $ and $ R $ light curves obtained at the same epoch and reported in \citet{Davanzo} (see Tab. \ref{phot}). 
\subsection{2009 Optical light curves}
\begin{figure}
\begin{center}
\includegraphics[scale=0.3]{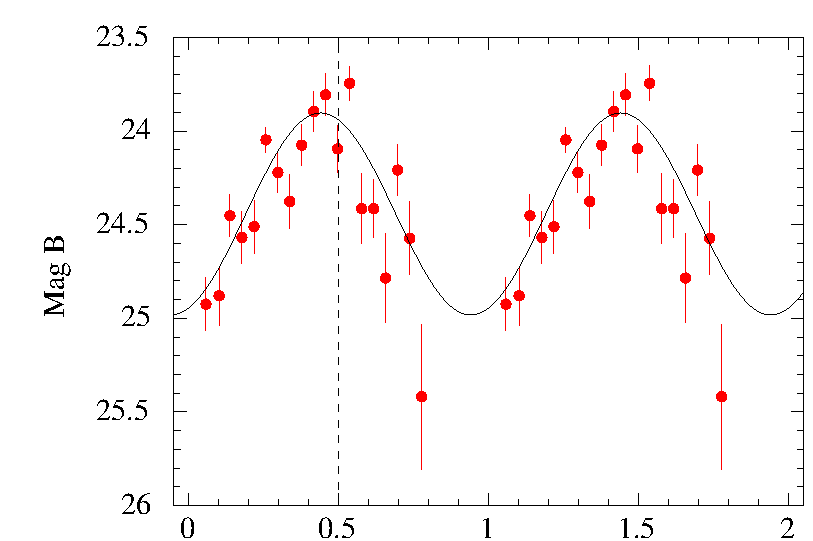}
   \includegraphics[scale=0.3]{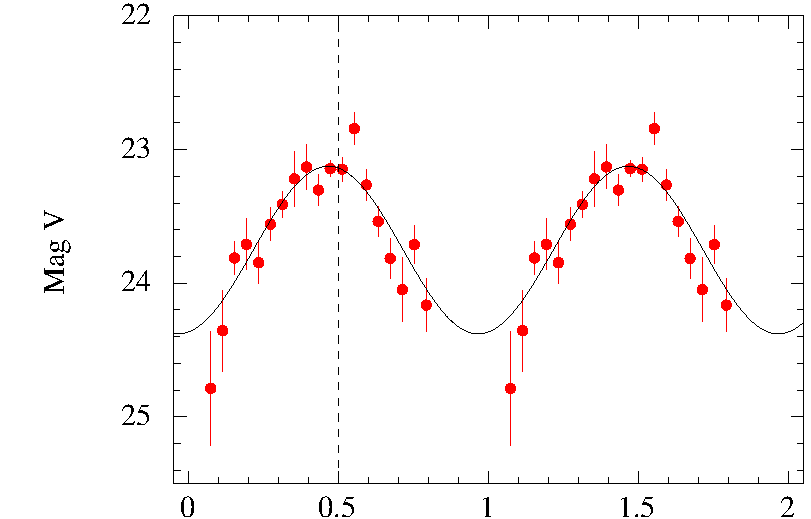}
   \includegraphics[scale=0.3]{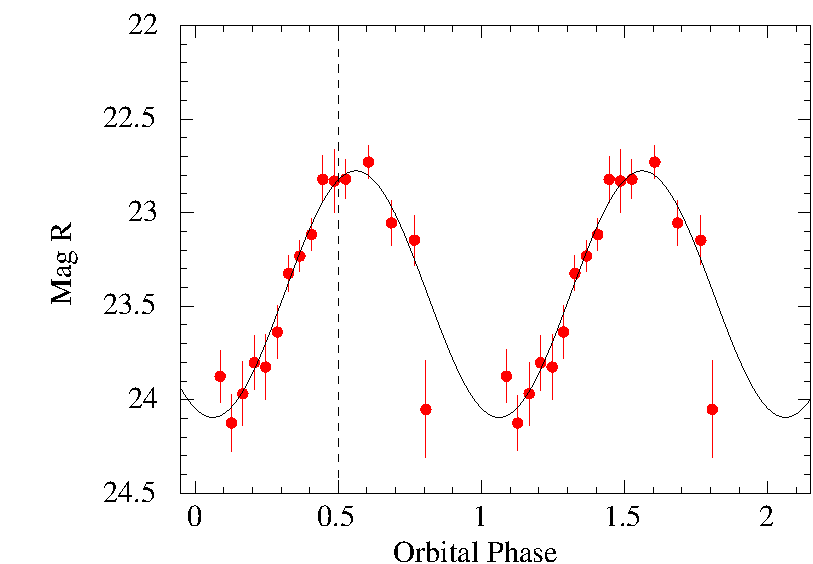}
   
   \caption{$B$, $V$, $R$ light curves for the system XTE J1814-338 for the 2009 observation. Phase 0 corresponds to the inferior conjunction of the companion. We calculated
             the phases based on the X-rays ephemeridis of \citet{Papitto}. Two periods of the system are drawn for clarity.}
\label{lc}
\end{center}
  
   \end{figure}

The optical counterpart of XTE J1814-338 (\citealt{Krauss2005}; \citealt{Davanzo}) is well detected in all our $ B $, $ V $, and $ R $ frames obtained under seeing $ \lesssim 1'' $ (see sect. 2). The optical phase resolved $BVR$ light curves are shown in Fig. \ref{lc}. The source in 2009 displays a clear sinusoidal variability, modulated at the $ 4.3  $ hours orbital period with a $ 0.5-0.7 $ mag  semi-amplitude (Table \ref{phot}). The $ B $ and $ V $ band light curves show a single minimum around phase 0.0 and a maximum consistent with phase 0.5 (based on the precise X-ray ephemerides by \citealt{Papitto}; see Table \ref{phot}). 

The same behaviour during quiescence was observed on May 2004 \citep{Davanzo}, suggesting that the compact object is heating the companion star. Such orbital modulated variability is also reminiscent of the quiescent optical light curves of other AMXPs (\citealt{Homer}; \citealt{Campana04}; \citealt{DavanzoIGR}; \citealt{Deloye}; \citealt{Wang09}; \citealt{Wang2013}) as well as recycled millisecond pulsars with irradiated companions (see, e.g., \citealt{Breton2013}).

Firstly we note that there seems to be an increasing trend of the light curves semi-amplitude with wavelength (see Tab. \ref{phot}), whereby the B-V colour is larger at inferior conjunction of the companion than at superior conjunction. Although such trend is affected by relatively large uncertainties, we note that it is opposite to what was observed during the 2004 observations (\citealt{Davanzo}) and to what one would expect from an irradiated companion since the heated side generally produces larger modulation amplitudes in the bluer part of the spectrum. 
A possible explanation for this puzzling behaviour is that some further component is contributing to the optical emission together with the companion star. In particular, the presence of a residual accretion disc during the quiescent state could explain our results. Since the \textit{RXTE/All Sky Monitor} light curve of XTE J1814-338 reveals no outburst between 2004 and 2009, it is possible that the accretion disc around the compact object grew during this period, thus increasing its relative contribution to the total quiescent optical emission of the system.
The disc is expected to be a hot structure with an emission peak at lower wavelengths with respect to the companion star, which peaks around the $ V $ band ($ \sim \rm 5500 \rm \,\, \AA $, \citealt{Davanzo}). The $B$ and $V$ band light curves could be for this reason more perturbed by the presence of the disc than the $R$ band one. Unless there are disc features displaying orbital variability (e.g. an eclipsed hot spot), its contribution to the total flux will be a constant value. As a result, the amplitude of the light curves will be attenuated. This effect is expected to be more important in the bluer bands since the disc spectrum should peak at a higher temperature than the companion.

    \begin{table}
\caption{Results of the photometry of XTE J1814-338 compared with the results of the observational campaign carried out on 2004 (\citealt{Davanzo}). The magnitude values are not corrected for reddening, whose parameters are reported in the last column. The 2004 $ I $ band results are obtained in this paper (text for details). The errors are indicated with a $90 \%$ c.l. }             
\label{phot}      
\centering                          
\begin{tabular}{c c c c c}        
\hline\hline                 
Filter  & Semi-amplitude & Mean & Maximum & $A_{\lambda}$\\    
        & (mag)	        &   Magnitude              & (phase)  &	\\ 
\hline                        
\multicolumn{5}{|c|}{2004 data \citep{Davanzo} }                     \\ \hline
   $V$ &$0.52 \pm 0.08$ & $23.29 \pm 0.04$ & $0.48 \pm 0.02$ & $0.87 \pm 0.03$ \\
   $R$ &$0.32 \pm 0.08$ & $22.52 \pm 0.03$ & $0.43 \pm 0.02$ & $0.66 \pm 0.06$ \\
   $I$ &$0.33 \pm 0.10$ & $21.79 \pm 0.04$ & $0.44 \pm 0.02$ & $0.50 \pm 0.03$ \\
\hline                                   
\multicolumn{5}{|c|}{2009 data (this paper)}                               \\ \hline
   $B$ &$0.52 \pm 0.10$ & $24.44 \pm 0.07$ & $0.44 \pm 0.02$ & $1.17 \pm 0.03$ \\      
   $V$ &$0.63 \pm 0.10$ & $23.75 \pm 0.08$ & $0.47 \pm 0.02$ & $0.87 \pm 0.03$ \\
   $R$ &$0.66 \pm 0.09$ & $23.43 \pm 0.05$ & $0.56 \pm 0.07$ & $0.66 \pm 0.06$ \\
\hline                                   
\end{tabular}
\end{table}

\noindent Furthermore, from our photometry we obtain unabsorbed mean colours $(B-V)=0.39 \pm 0.11 \, \rm mag$ and $(V-R)=0.11 \pm 0.12 \, \rm mag$ (computed assuming $E(B-V)=0.29$; \citealt{Davanzo}). If considered as the colours of a main sequence star (as the companion star of XTE J1814-338 probably is; \citealt{Krauss2005}; \citealt{Davanzo}), these values translates into inconsistent surface temperatures (between 6250 K and 7000 K from $(B-V)$ and between 11400 K and 7300 K from $(V-R)$; \citealt{Allen}), providing a further indication of a two-component emission. 
\subsection{Comparison between 2004 and 2009}
\begin{figure}
\begin{center}
\includegraphics[scale=0.3]{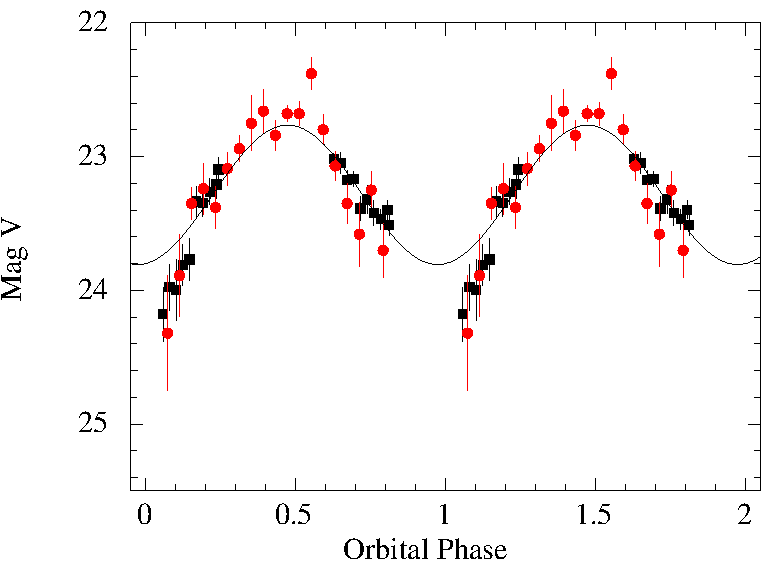}   
\caption{$V$ light curve taken in 2009 (dots, this work) and the one taken in 2004 (squares, \citealt{Davanzo}). Both curves are normalized to the mean 2009 magnitude.}
\label{2004_2009}
\end{center}
\end{figure}

\begin{figure}
\begin{center}
\includegraphics[scale=0.3]{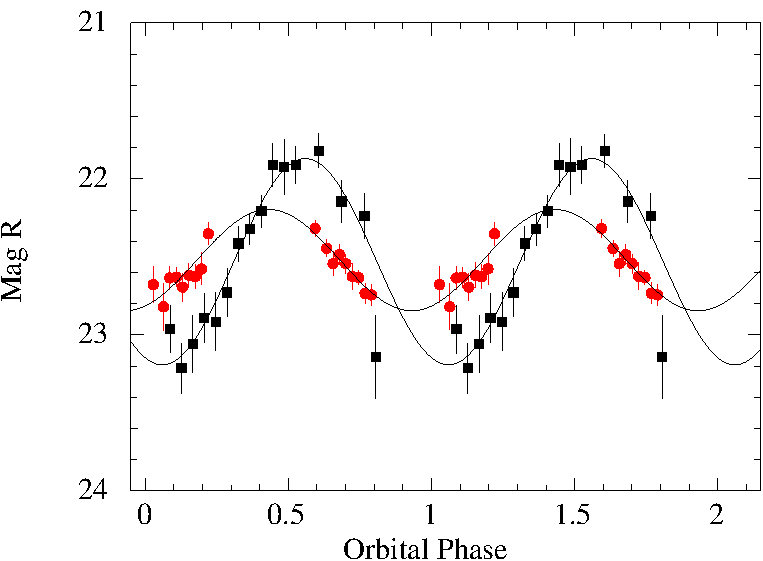}  
\caption{$R$ light curves of \citet{Davanzo} (dots) and the one referred to the 2009 data (squares). Both curves are normalized to the mean 2009 magnitude. }
\label{R20042009}
\end{center}
\end{figure}

We note that the system was significantly fainter in the optical in 2009 than in 2004. Comparing the two datasets (Tab. \ref{phot}) we find that the mean $V$ and $R$ band luminosities decreased by $0.46 \pm 0.08 \, \rm mag$ and $0.91 \pm 0.06 \, \rm mag$, respectively. The combined $V$ and $R$ band light curves obtained from both datasets, rescaled to the same mean magnitude, are reported in Fig. \ref{2004_2009} and Fig. \ref{R20042009}. While the combined $V $ band curve can be satisfactorly fitted with a single sinusoid plus constant model, this is clearly not possible with the $R$ band data  (Fig. \ref{R20042009}), because of a significant phase offset.

This is unexpected if the irradiated star is responsible for the entire quiescent emission of the system. In this case we would expect to see all light curves, in the same band, peaking at the same phase, since in that scenario all the light curves should peak at phase 0.5 (i.e. superior conjunction of the companion), without any significant offset. 
This suggests that part of the modulation arises from another location, such as a disc asymmetry (e.g. a hot spot) whose periodicity is different from the orbital period of the companion. However, this scenario would require similar phase shifts to be observed in light curves from other bands as well, which is not the case (Fig. \ref{R20042009}). On the other hand, alternative explanations, like superhumps due to disc precession, are most likely expected to occur during outburst phases. 

Offsets with respect to phase 0.5 are also observed in the \textit{B} and \textit{I} band light curve maxima. However, for such bands we have data covering only a single epoch (2004 for the \textit{I} band, 2009 for the \textit{B} band) which prevent us from making a comparison (as we did for the \textit{V} and \textit{R} band) and speculate on the possible physical origin of this offset.
\subsection{Fitting with {\tt ICARUS}}\label{ICARUS}

 \begin{figure*}[!t]
\begin{center}
\includegraphics[scale=0.5]{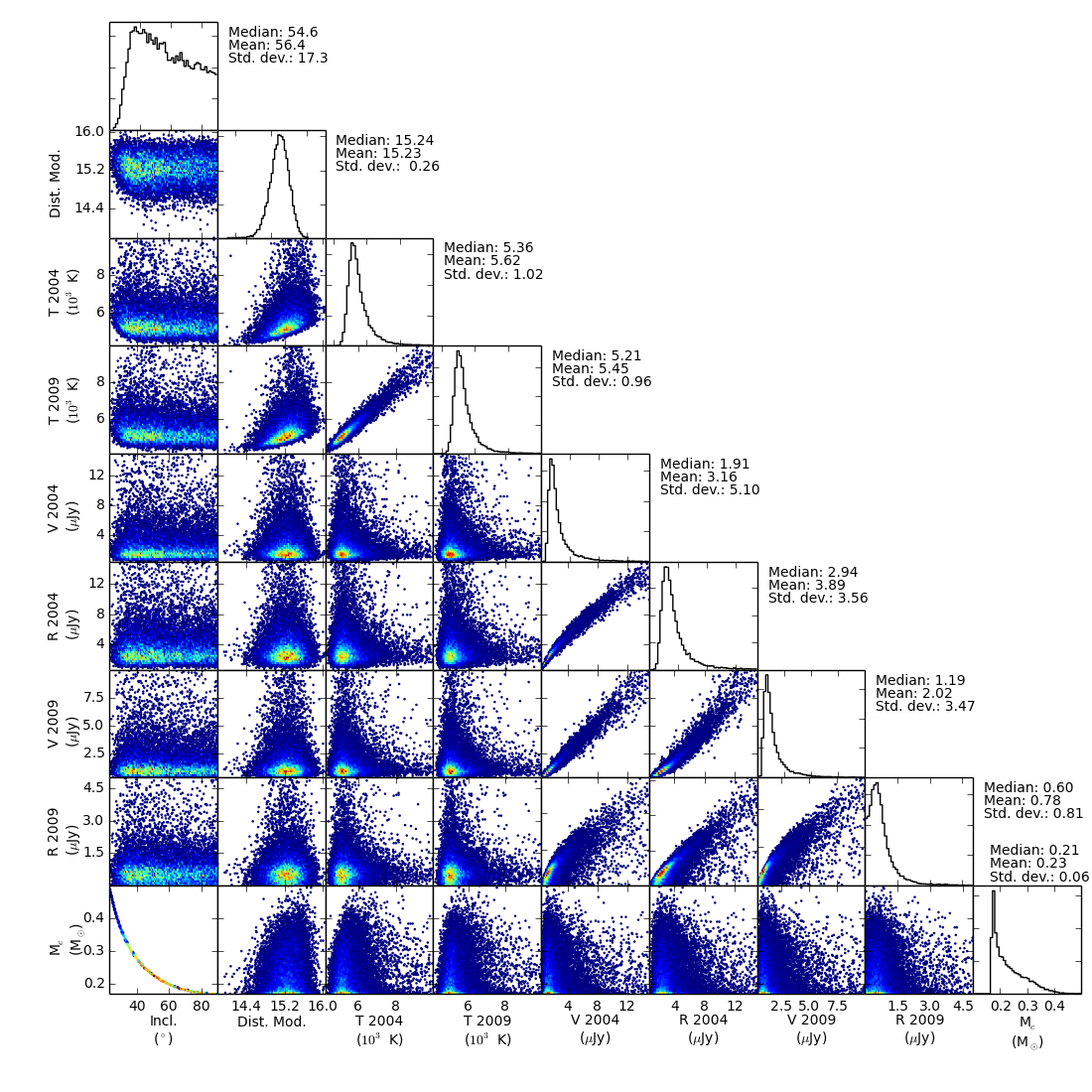}   
\caption{One- and two-dimensional posterior distributions of key parameters obtained from our light curve fitting. The $R$ and $V$ denote the disc fluxes ($\rm \mu Jy$) in their respective bands in 2004 and 2009, $T$ refers to the day side temperatures and $ M_{\rm comp} $ is the companion star mass. For these fits we adopted $M_{\rm ns} = 1.4 \, M_\odot$.}
\label{icarus}
\end{center}
\end{figure*}
In order to model the
XTE J1814-338 optical light curves with a self-consistent physical approach, we fitted the data using the Icarus\footnote{Freely available at https://github.com/bretonr/Icarus} light curve modelling suite for irradiated binary companions \citep{Breton2012}. This model comprises several free parameters, which are:
\begin{itemize}
\item orbital inclination
\item distance modulus
\item extinction in J band 
\item day side companion temperature in 2004
\item day side companion temperature in 2009
\item disc flux in V band in 2004
\item disc flux in R band in 2004
\item disc flux in V band in 2009
\item disc flux in R band in 2009
\item disc flux in B band in 2009
\end{itemize} 

Given a mass for the pulsar, one can use the pulsar mass function derived from the X-ray timing to determine the companion mass. Then, the back side temperature of the companion is determined using a mass-temperature relationship for low-mass brown dwarfs \citep{Deloye}.

We performed the light curve fitting using a Markov chain Monte Carlo procedure similar to that reported in \citet{Wang2013}. We used a normal prior on the distance of $8.0 (M_{\rm ns}/1.4 M_\odot)$ with 20$\%$ uncertainty, which corresponds to the value derived from type I bursts by \citet{Strohmayer2003}. For the reddening, we used $A_{J} = 0.253 \pm 0.078$, which is the value reported in \citet{Davanzo} with conservative uncertainties inflated by a factor of three.

Assuming a neutron star mass of 1.4
$M_{\odot}$, the code provides a satisfactory fit (Fig. \ref{icarus}) with a distance modulus of 15.2 (which corresponds to a system distance of $ \sim $ 11 kpc, within $ 2\sigma $ from the estimated distance from type I bursts), a mean companion day side temperature of $\sim 5500$ $K$ and a companion mass of $(0.23 \pm 0.06) \,
M_{\odot}$ (in agreement with the estimates reported in \citealt{Davanzo}). We do not find any evidence for a variation of the day side temperature between 2004 and 2009. However, a clear change in the disc flux is found during this period, with a decrease by a factor of 1.6 and 5 in the $V$ and $R$ band, respectively. This shows
that the disc became fainter and bluer. The orbital inclination, on the other hand, is basically unconstrained. Finally, we note that similar results
are found assuming a neutron star mass of 2.0 $M_{\odot}$. 
 
\section{Discussion}
A comparison and joint analysis of multiband optical data of XTE J1814-338 over 5 years shows that the quiescent optical emission of the system is likely due to the
combination of at least two different components, namely an irradiated companion star and a disc.
\subsection{The companion star}
The combined fit of the optical light curves
(sect. \ref{ICARUS}) yields a companion star mass of $(0.23 \pm 0.06) \, M_{\odot}$, a value
typical of main-sequence $M-$type stars. On the other hand, the inferred companion
day-side surface temperature of $ \sim $5500 $K$ (found to be constant
during both the 2004 and 2009 obbservations) would imply an earlier spectral type (G-K) for a main-sequence star. Such apparent discrepancy was already pointed out and
discussed in \citet{Davanzo} as the indirect evidence for a companion star heated
by the relativistic particle wind of the neutron star, re-activated as a millisecond radio
pulsar during the system quiescence. These results are in agreement with the so-called
``recycling scenario'', in which neutron stars from AMXP systems switch between the state of accretion-powered pulsar to rotation-powered pulsar when the accretion stops (\citealt{Alpar82}; \citealt{Stella}; \citealt{Campana98}; \citealt{Burderi}; \citealt{Patruno}). A similar evidence was found by \citet{Wang2013} from the analysis of time-resolved optical imaging of SAX J1808.4-3658, the
prototypical AMXP, collected during outburst and quiescence.

The absence of significant day-side temperature variation in the XTE J1814-338 companion during the five-year period suggests that a few months after the end of the 2003 outburst, the neutron star was already back to its rotation-powered state (see also \citealt{Papitto2013}).
Assuming that irradiation simply acts to increase the companion's surface temperature, one can write the irradiation luminosity as $L_{\rm irr} = \sigma_{\rm SB} (T_{\rm day}^4 - T_{\rm night}^4)$, where $\sigma_{\rm SB}$ is the Stefan-Boltzmann constant. If the irradiating flux is provided by the re-activated millisecond pulsar through its relativistic wind, one can relate the irradiation luminosity to the spin-down energy as $L_{\rm sd} = 4 \pi a^2 L_{\rm irr} / \epsilon_{\rm irr}$. Here, $a$ is the orbital separation and $\epsilon_{\rm irr}$ is the irradiation efficiency, which can be ascribed to the albedo of the companion and the isotropy of the spin-down energy transported by the wind. \citealt{Breton2013} showed that $\epsilon_{\rm irr}$ for typical irradiated pulsar systems is in the range $0.1-0.3$. Hence, using the fitted temperature of the companion ($\sim\,5500$ and $3300\,K$ on the day and night side, respectively), we estimate the spin-down energy of the neutron star to be in the range $L_{\rm sd} \sim [6-17]\times 10^{34}\,\rm erg\, \rm s^{-1}$, which is typical for a $3.2\, \rm ms$ pulsar.

\subsection{The accretion disc}
While the companion star contribution to the overall quiescent optical emission remained constant, the disc component displayed a significant change during the five years of our
observations, both in terms of flux and colour. In particular, the observed spectrum
became bluer as the flux decreased from 2004 to 2009. A possible explanation of such long-term evolution will be discussed in the following section.

\begin{figure}
\begin{center}
\includegraphics[scale=0.29]{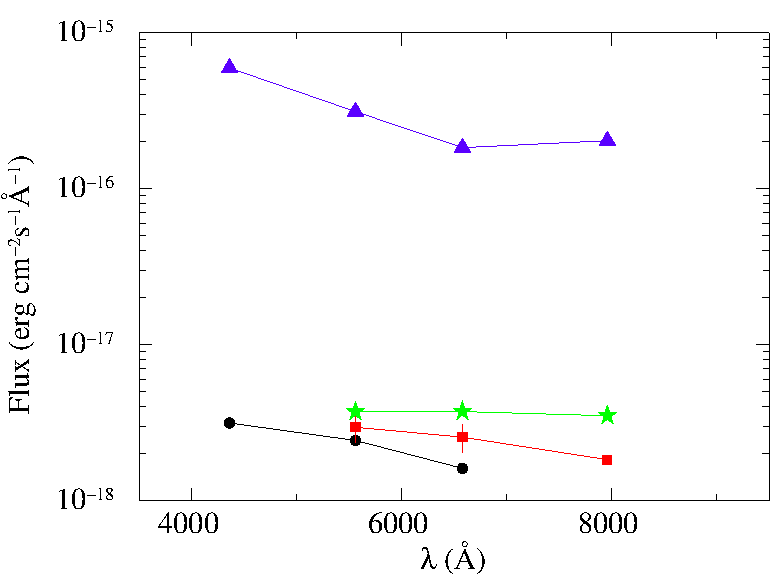}   
\caption{Spectral energy distribution of XTE J1814-338 at four different epochs: 2003 (triangles, \citealt{Krauss2005}), 2004 (stars, \citealt{Davanzo}), 2007 (squares, \citealt{Davanzo}) and 2009 (dots, this paper).}
\label{flux_lambda}
\end{center}
\end{figure}

\subsection{A jet as a possible third component}
XTE J1814-338 is one of the three
AMXPs (together with SAX J1808.4-3658 and XTE J0929-314) for which a flux excess has been
observed in the near-IR with respect to the extrapolation of the UV-optical flux. Such excess
has been observed for all three systems during the outburst phase at a wavelength
corresponding to the $I$ band and has been interpreted as the contribution of a
synchrotron emitting region, such as a synchroton jet (\citealt{Krauss2005}; \citealt{Wang2001}; \citealt{Giles2005}; \citealt{Russell2007}). In Fig. \ref{flux_lambda} we plot the evolution of the optical spectral energy
distribution (SED) of XTE J1814-338 from the 2003 outburst to quiescence. The data plotted in Fig. \ref{flux_lambda} are taken from \citet{Krauss2005} and from Table \ref{phot} of this paper. We also added the corresponding $VRI$ dereddened
fluxes derived from the low-resolution optical spectrum of XTE J1814-338 obtained in 2007 during
quiescence \citep{Davanzo}. Despite the dramatic difference in terms of fluxes, the $(R-I)$ colour is almost constant in the June 2003 and May 2004
SED (\citealt{Krauss2005} found $ 0.85 \pm 0.06$ and \citealt{Davanzo} $ 0.73 \pm 0.05$), suggesting that the extra component responsible of the $I $ band excess observed by \citet{Krauss2005} during the outburst might still be present at the beginning of the quiescent phase. Such a component seems to be absent (or significantly fainter) in the following SEDs obtained in 2007 and 2009 during quiescence. The presence of a transient near-IR component due to synchrotron emission may provide an explanation to the large colour change observed in the quiescent optical SEDs. Indeed we observe the spectrum becoming fainter and bluer. This could be explained if the jet
component is progressively fading or its break frequency is moving toward longer
wavelengths (i.e. lowering its contribution to the optical emission). Although jet emission in X-ray binaries is typically observed during active phases (\citealt{Fender01}; \citealt{Russell2013}), we note that it has recently been suggested to contribute to the quiescent optical emission from the black hole candidate Swift J1357.2-0933 \citep{Shahbaz2013}. However we are missing observations in the mid-IR and radio bands, where the possible jet contribution is expected to dominate \citep{Fender01}, to corroborate our claim. 

An alternative explanation for the observed spectral evolution might be that between 2004 and 2009 the disc underwent a mass-transfer instability episode during which material from the outer part of the disc was transported inwards as a result of a viscous change. This effect could induce a fading of the outer disc region, whose emission is mostly driven by irradiation from the compact object ($ T_{\rm irr}\propto R^{-0.5} $ whereas $ T_{\rm visc}\propto R^{-0.75}$, \citealt{Hynes2010}). A similar explanation was proposed by \citet{Giles2005} to
explain the spectral evolution observed during the outburst of the AMXP XTE J0929-314.
\subsection{The phase offset}
The main contribution to the quiescent optical emission observed in the 2009 data comes from the companion star, irradiated by the neutron star. Considering a companion star with a mass of 0.23 $ M_{\odot} $ filling its Roche lobe, we find that the whole star extends $ \sim 0.07 $, consistent with the 0.06 observed phase-offset.

Since the irradiation from the compact object should be symmetric, we hypothesize the presence of an additional component to the irradiation, likely a hot spot located in the accretion disc. Indeed, a consequence of the ``hot spot irradiation'' is that the inner face of the companion star would be be asymmetrically illuminated, causing the light curve maximum to shift from phase $\sim 0.5$ (as expected in the case of irradiation from the compact object) to phase $\sim 0.6$ (see Fig.~6 of \citealt{Beuermann90}). We note that evidence for asymmetrically irradiated companion stars caused by hot spot irradiation during quiescence in LMXB systems and cataclysmic variables has already been found (\citealt{Davanzo2006}, \citealt{Beuermann90}). 

\section{Conclusions}
In this work we presented the results of a phase-resolved multiband study of the quiescent optical counterpart of the AMXP system XTE J1814-338 obtained in 2009. Similar observations of the same source have been conducted by our team five years before at an earlier stage of the quiescent phase. The source experienced no outburst during this time period. The main results we obtained are summarized below.  
\begin{itemize}
\item We have a clear detection of the quiescent optical counterpart of the system. The multiband ({\it B, V, R}) phase-resolved light curves show variability modulated at the 4.3 hr orbital period.
\item There is evidence of a two-component emission both in the 2004 and the 2009 observations, namely an accretion disc and a heated companion star.
\item MCMC fitting of the multiband light curves enabled us to constrain some system parameters as the distance, the companion star's mass and surface temperature.
\item Comparing the 2004 and 2009 datasets we find that the system is on average fainter and bluer in the optical during the latter period. This is linked to the presence of the disc emission component, that changed during the five-year period, while the companion contribution remained constant. A suitable interpretation of the observed flux and spectrum is the presence of a third component in the 2004 emission, such as a jet, which is supported by the $ I $ band excess observed for the system during the outburst in 2003 by \citet{Krauss2005}. This component would have disappeard or severally weakened from 2004 to 2009, thus explaining the behaviour of the continuum component of the optical emission.
Alternatively, a mass transfer instability due to viscous forces could explain the observed colour evolution.
\item A phase offset is observed between the 2004 and 2009 $R$ band light curves (but not in the corresponding $V$ band data) suggesting that the companion star might be asymmetrically irradiated. We speculate that a combination of heating effects on the companion star from both the compact object and a hot spot in the disc could explain the observed phase offset.
\end{itemize}

\begin{acknowledgements}
MCB acknowledges the INAF-Osservatorio Astronomico di Brera for kind hospitality during her bachelor thesis.
SC and PDA acknowledge the Italian Space Agency (ASI) for financial support through the ASI-INAF contract I/004/11/1.
TMD acknowledges funding via an EU Marie Curie Intra-European Fellowship under contract no. 2011-301355. 
RPB was funded by a European
Research Council Advanced Grant 267697 ``4 Pi Sky: Extreme Astrophysics with
Revolutionary Radio Telescopes''.
\end{acknowledgements}



\end{document}